\documentclass[journal,11pt]{IEEEtran}
\usepackage{cite}
\usepackage[pdftex]{graphicx}
\usepackage{amsmath}
\usepackage{multirow}
\usepackage{algorithmic}
\usepackage{array}
\usepackage{url}
\ifCLASSOPTIONcompsoc
  \usepackage[caption=false,font=normalsize,labelfont=sf,textfont=sf]{subfig}
\else
  \usepackage[caption=false,font=footnotesize]{subfig}
\fi
\begin{document}

\title{Smart Load Node for Non--Smart Load\\ 
under Smart Grid Paradigm:\\
A New Home Energy Management System}
\author{Shashank~Singh,~\IEEEmembership{Student~Member,~IEEE},
		Amit Roy,
        and Selvan~M.P.,~\IEEEmembership{Member,~IEEE}
}

\markboth{Published: IEEE Consumer Electronics Magazine. INDIAN PATENT APPLICATION, APP NO: 201741039083,~2017-NOV-2}%
{Shell \MakeLowercase{\textit{et al.}}: Bare Demo of IEEEtran.cls for IEEE Journals}

\maketitle

\begin{abstract}
This article presents a novel approach for efficient operation of non--smart household appliances under smart grid environment using the proposed smart load node (SLN). In real world scenario, there are so many non--smart loads currently in use and embedding appliance specific intelligence into them to make them as smart loads will be more expensive compared to the proposed SLN, which is a common solution for all types of non--smart loads. This makes the proposed low--cost SLN, which neither requires any infrastructural change in the electrical wiring of a house nor any constructional change in home appliances at the manufacturing stage and at the consumer end, as a feasible solution for intelligent operation of non--smart home appliances under smart grid environment. The SLNs, which are placed in a home like distributed wireless sensor nodes, form a home area network (HAN). The HAN includes a load management unit (LMU) which acts as master for all distributed SLNs. Wi--Fi is chosen as a medium of communication in the HAN. The LMU incorporates load management algorithm which is written in Python script.  
\end{abstract}

\IEEEpeerreviewmaketitle

\section{Smart Load Node: What Is It And Why Is IT Needed}
\IEEEPARstart{T}{he} smart grid is an intelligent grid which is featured with smart metering technologies, modern power converters, rapid communication infrastructure, automation and consumer participation that leads to efficient and reliable operation of power system \cite{gungor2011}. Smart grid presents many opportunities and challenges in the field of electrical sciences \cite{yu2011}. Development of home area network (HAN) for intelligent scheduling of the operation of non-smart loads is one of the aforementioned challenges. A home area network is the lowest layer of smart grid where electrical parameters of various loads are recorded and communicated to a processing unit \cite{noorwali2016}. The processing unit of a HAN is referred to as load management unit (LMU) in this article. 

\begin{figure}[htbp]
\centering
\includegraphics[width = \columnwidth]{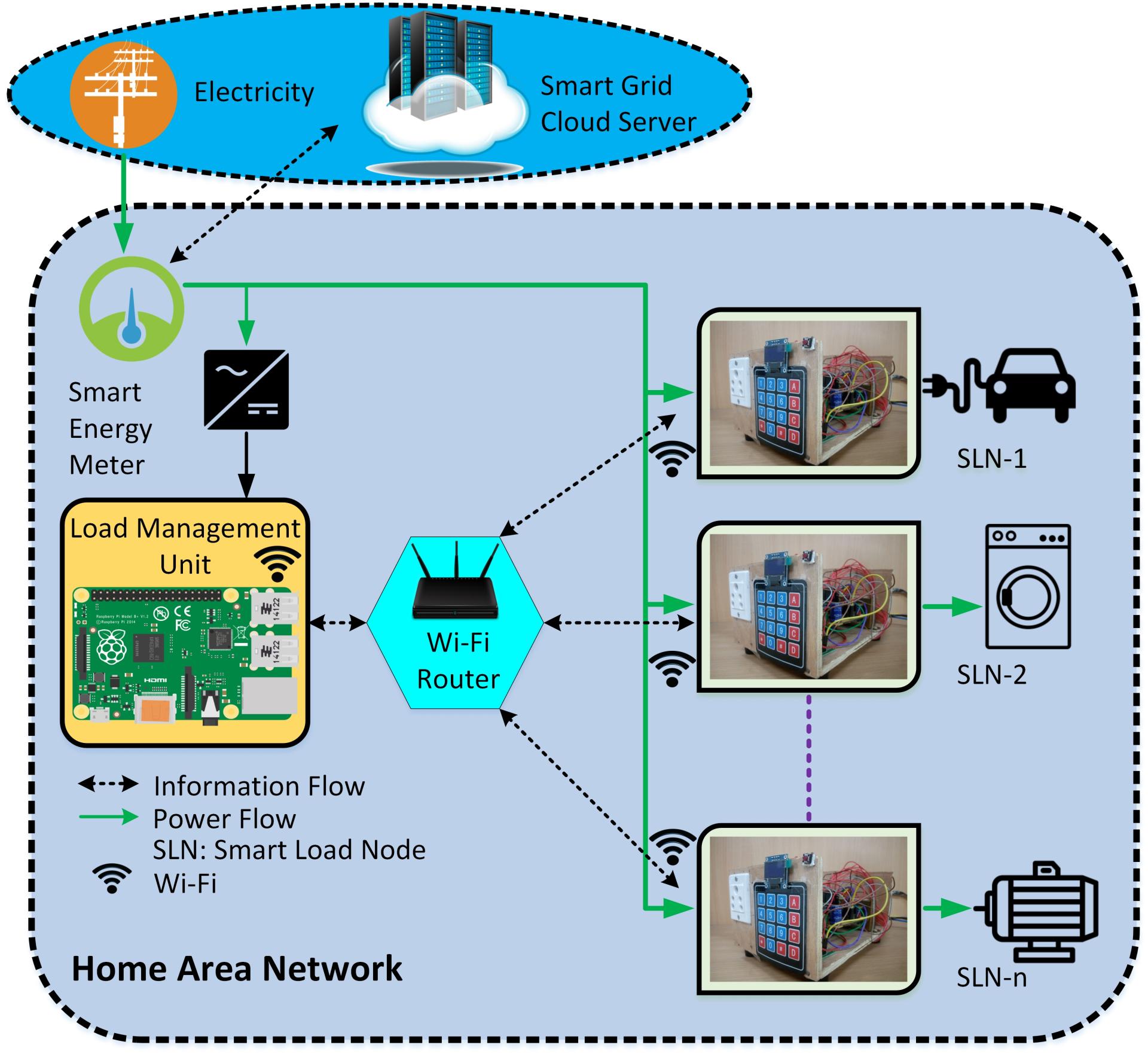}
\caption{LMU and SLNs connected via Wi--Fi link}
\label{LMU}
\end{figure}

The continuous increase in the consumption of electrical energy in residential buildings has received more attention towards the development of home energy management systems (HEMSs). It is evident from past studies that consumers are interested in money--saving through time--of--use pricing and price--signals \cite{chang2012}. Demand response (DR) and HEMS are described in \cite{yuan2012}. It conveys that HEMSs along with smart energy meters \cite{singh2017} will bring DR on a wide scale. IEEE 802.15.4 and ZigBee based HAN is developed by Han and Lim, which provides light control and multi-sensing design. Their architecture deals with the implementation of dynamic sensor network and control in a house, which requires significant modification in electrical infrastructure of a house \cite{han2010}, \cite{lim2010}. Two approaches have been suggested to establish communication between the home appliances and LMU \cite{namboodiri2014}. The first approach is to make the appliance smart by incorporating communication and information processing facility. At present, appliances that are already in use at the consumer end do not have such facility, which is a bottleneck for this option. Such appliances manufactured without any information and communication processing features are referred to as non--smart loads. The second suggested approach is to embed intelligence in power outlets, which requires modifications in the electrical wiring infrastructure of a built house.  

In this article, a smart load node (SLN) is proposed as third approach to develop a HAN for efficient scheduling of home appliances by incorporating DR programs \cite{vardakas2015}, \cite{palensky2011}. SLN is meant to be connected between the existing power outlet and the load. Each SLN has a user interface that helps the consumer to configure the operation of a load. Upon configuration by the consumer, SLN sends load parameters to LMU, a dedicated single--board computer that runs a load management algorithm for intelligent scheduling of loads to participate in DR programs. The flow of information and power in the proposed HAN is depicted in Figure 1. The novel aspects of SLN are:-

\begin{itemize}
\item Requires no modifications in the existing electrical infrastructure of a house.
\item Requires no modifications in the household appliances at the manufacturing stage and at the consumer end.
\item A Common solution for all types of non--smart loads.
\item Capable of incorporating various non--smart schedulable loads in any kind of DR algorithm.
\item Mobile in nature. In case of change of house, same HAN comprising of SLNs can be used without any additional expenses on components.  
\item Acts as wireless power, power factor, voltage and current meter.
\end{itemize}

\section{Classification of Residential Loads}
Residential loads are grouped into schedulable and non--schedulable load categories \cite{arun2018}. Schedulable loads are further classified into two categories namely interruptible and schedulable loads (ISLs) and non--interruptible and schedulable loads (NISLs). Non--schedulable loads are hereafter referred to as non--interruptible and non--schedulable loads (NINSLs). Essential loads such as light, fan, entertainment loads (TV, DVD player, personal computer, printer, mobile charger and laptop charger) and kitchen loads (cooking stove, mixer and toaster) are classified as NINSLs because the operating time of these loads completely depends upon the desire and comfort of the user. Loads which cannot be interrupted during their operation but are flexible with the time of use are classified as NISLs, e.g., clothes dryer, clothes washer, and mixer grinder. ISLs can operate in a continuous or discontinuous manner in the user--defined schedule. Plug--in hybrid electric vehicle (PHEV), dishwasher and well--pump fall under this category.

\section{Design \& Features of SLN}
The major elements of SLN are Arduino UNO--R3 board (Arduino), voltage and current sensing units, ESP8266 Wi--Fi transceiver module, relay, 0.96" OLED display, and a matrix keypad. The block diagram of SLN is depicted in Figure 2. 

\begin{figure}[htbp]
\centering
\includegraphics[width = \columnwidth]{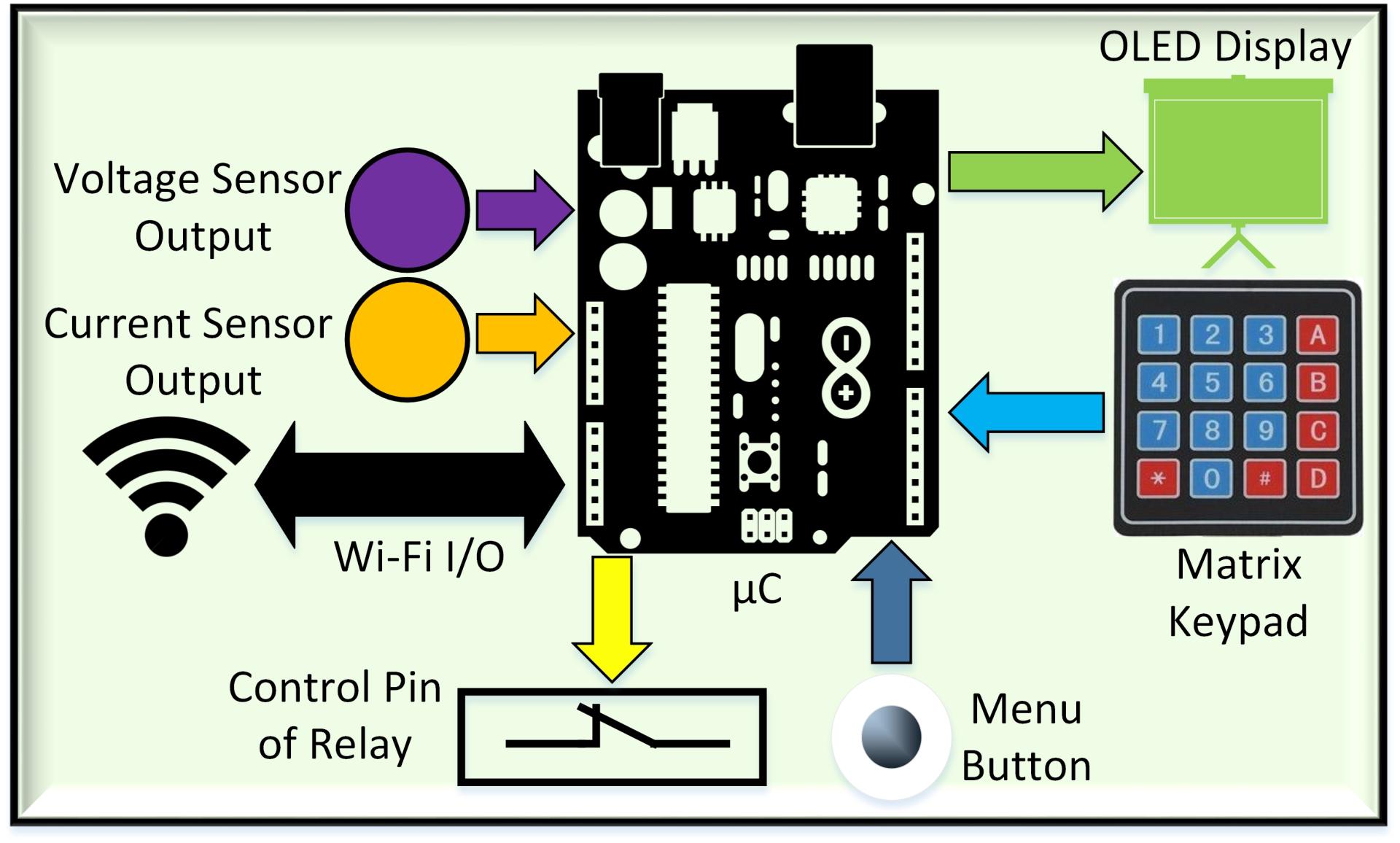}
\caption{Arduino with multiple input/output devices}
\label{SLN Block}
\end{figure}

The load must be connected to the wall socket through the SLN. The supply voltage, tapped from the wall socket is stepped down by a transformer and added with a DC offset before applying to A0 pin of Arduino. ACS712 current sensor is connected in series with the phase of AC supply. The output of the current sensor is applied to A2 pin of Arduino. The acquired voltage and current data from A0 and A2 pins are processed to calculate the electrical parameters of a load. A dedicated transformer serves the purpose of supplying power to various elements present in SLN. The Wi--Fi transceiver of SLN has a static IP address. It is connected to Arduino through the serial port to send load parameters and load configurations set by the user (logged using matrix keypad) to LMU and to receive instructions for the operation of a load. One of the developed SLNs shown in Figure 3a, has three sub-displays in one OLED display. Displayed information in Figure 3b is the default display of SLN. The configuration of load can be changed by pressing the menu button. Selection of node configuration (Figure 3c) leads consumer to data logging screen (Figure 3d).  Upon the data logging request, the consumer must provide start time ($\alpha$, starting interval), stop time ($\beta$, interval by which the load must complete its operation) and interval ($\gamma$, duration of operation of the load in minutes). The expression (1) conveys that upon selection of NINSL mode, SLN will not ask for any user input.

\begin{figure}[htbp]
\centering
\includegraphics[width = \columnwidth]{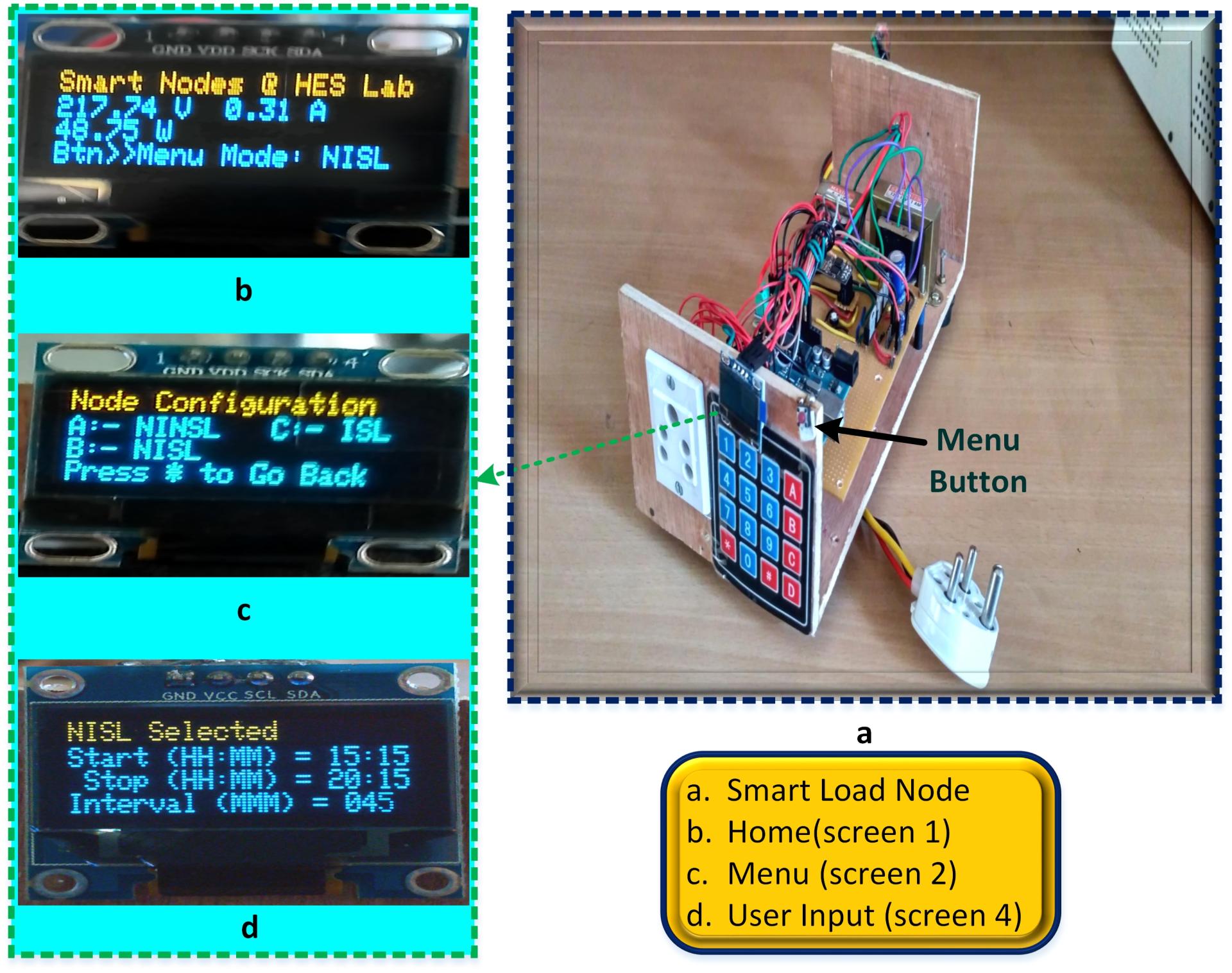}
\caption{Developed SLN depicted with its user interface}
\label{SLN}
\end{figure}
\begin{equation}
\left( \alpha ,\beta ,\gamma  \right)=\left\{ \begin{matrix}
   \left( \text{0,0,0} \right) & if  \\
   \left( \text{1,1,1} \right) & if  \\
   \left( \text{1,1,1} \right) & if  \\
\end{matrix} \right.\begin{matrix}
   NINSL  \\
   NISL  \\
   ISL  \\
\end{matrix}
\end{equation}

If the consumer chooses either of NISL and ISL mode then $\alpha$, $\beta$, and $\gamma$ must be provided as inputs. LMU issues ON and OFF command between $\alpha$ and $\beta$ intervals considering the operating duration $\gamma$ of load. In case of NISL mode, load must not be interrupted during its operation between $\alpha$ and $\beta$. Upon reception of ON command, NISL load will run uninterruptedly for a period of $\gamma$. In case of ISL mode, load may be interrupted during its operation through multiple ON/OFF command dictated by DR algorithm, however the sum of intermittent run time of load must be equal to $\gamma$. The concept of interval and referred DR algorithm is explained in section 5.

\section{Home Area Network Architecture}
LMU is a Raspberry Pi 3 Model B processor which executes a load management algorithm. LMU is connected to a smart energy meter (SEM) and SEM is connected to the smart grid via internet.  SEM receives values of parameters required to incorporate DR in load management algorithm and transfers those values to LMU. Based on the configuration set by the consumer, SLN formats the logged input data and sends it to LMU. Upon reception, LMU splits the data with the help of format specifiers, and stores the respective values separately in its memory. LMU is synchronized with GMT +5:30 (Indian Standard Time). The ON and OFF signals are generated by LMU which trigger the inbuilt relay of respective SLN.

\begin{figure*}[htb]
\centering
\includegraphics[scale=0.25]{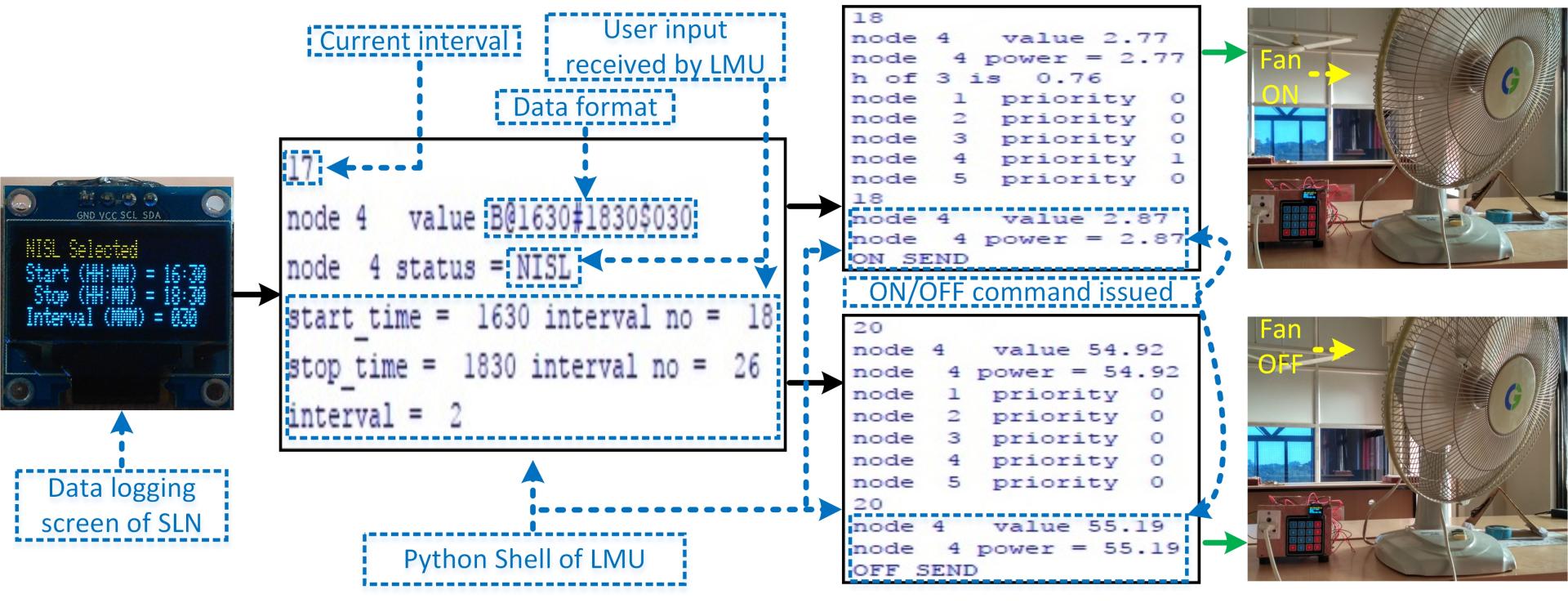}
\caption{Insight of HAN operation, from data logging to the operation of load}
\label{HAN}
\end{figure*}

Figure 4 depicts data processing and routing followed by load operation. Data reception by LMU and load triggering are evident from Python shell of LMU in HAN. The concept of load priority and interval of load operation is a subject of DR algorithm which is explained along with preceded terms in the next section.

\section{Case Study and Results}
It is impractical to request consumers to develop an efficient load scheduling algorithm to reduce their bills and help towards the healthy operation of grid. This problem leads to the development of automatic load scheduling algorithm \cite{liu2012,arun2017,chan2012,rodriguez2016} for a smart building \cite{mohanty2016}.  To demonstrate the capability of execution of DR program through proposed SLN, two case studies are performed. The first case study does not employ any load scheduling algorithm and the second case study employs one of the priority based load scheduling algorithms \cite{arun2017}. Figure 5 shows intermediate steps involved in the dynamic priority based load scheduling algorithm employed in this case study. It is assumed in this case study that utility is providing maximum demand limit (MDL) one day ahead via SEM. MDL is defined as the maximum permissible amount of power that can be drawn in that hour without penalty.

\begin{figure}[htbp]
\centering
\includegraphics[scale = 0.2]{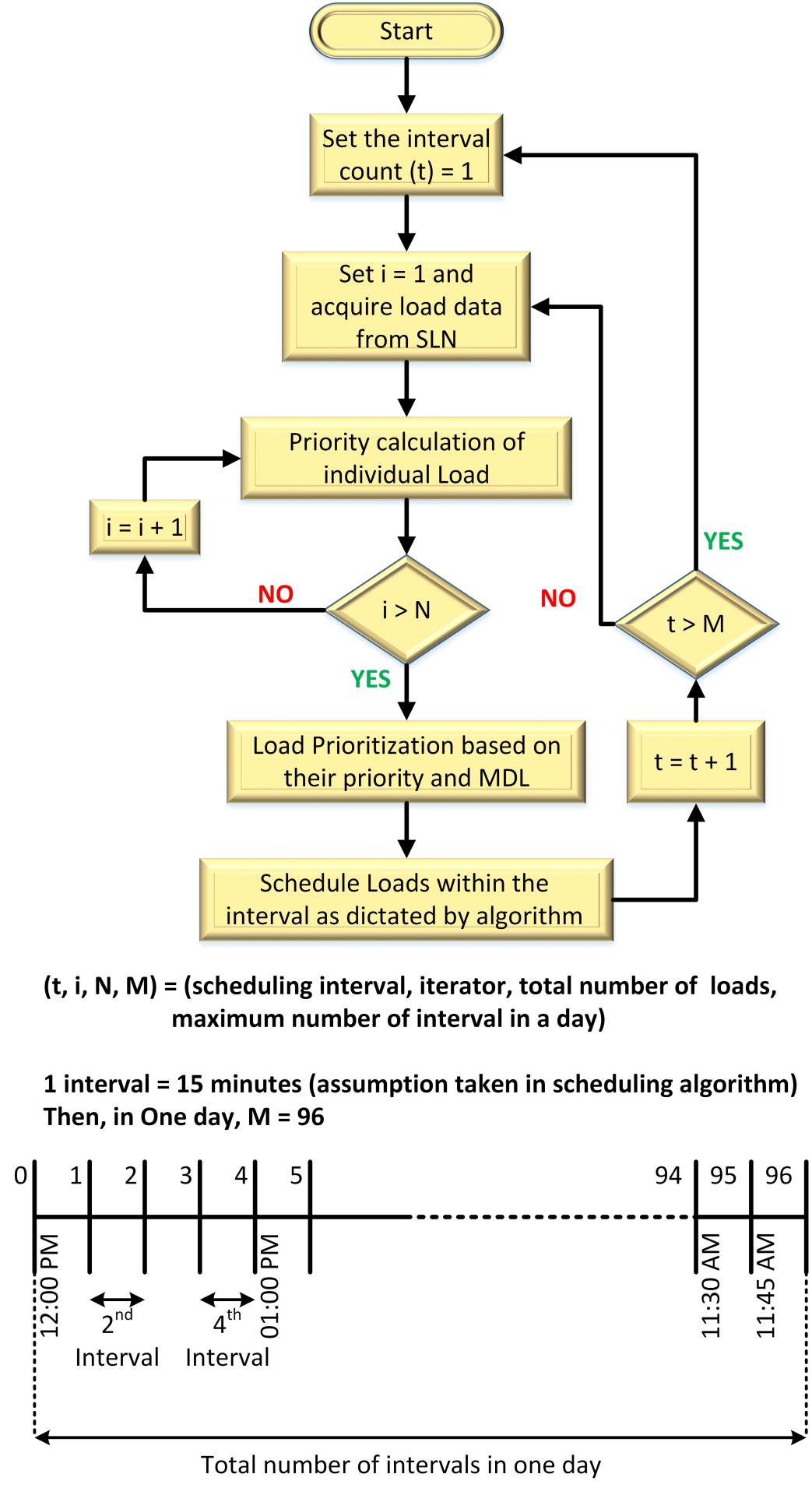}
\caption{Illustration of dynamic priority based load scheduling algorithm }
\label{Algo}
\end{figure}

\begin{figure}[htbp]
\centering
\subfloat[Load curve without any scheduling algorithm]{\includegraphics[width=\columnwidth]{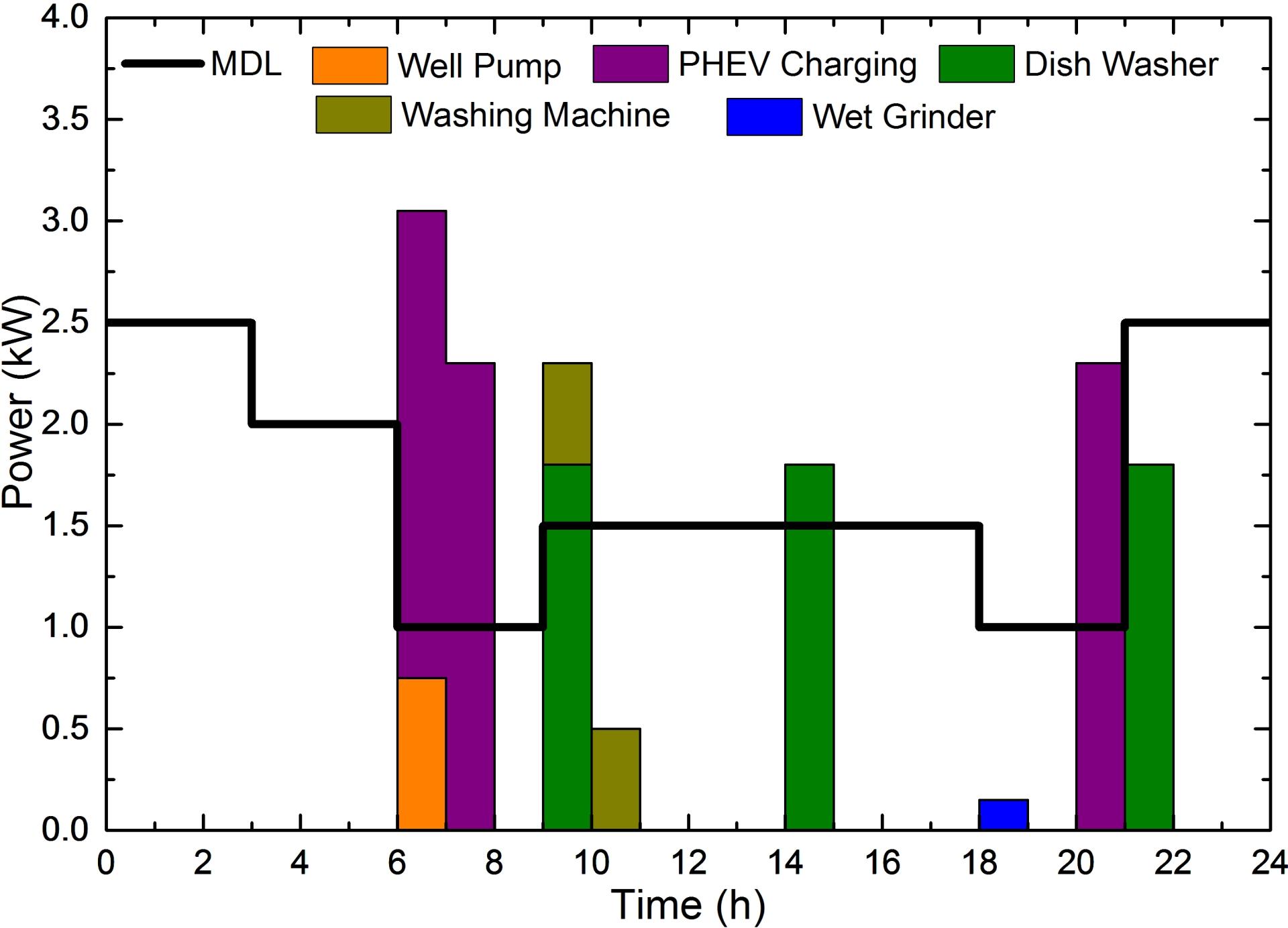}
\label{6a}}
\hfill
\subfloat[Load curve with scheduling algorithm]{\includegraphics[width=\columnwidth]{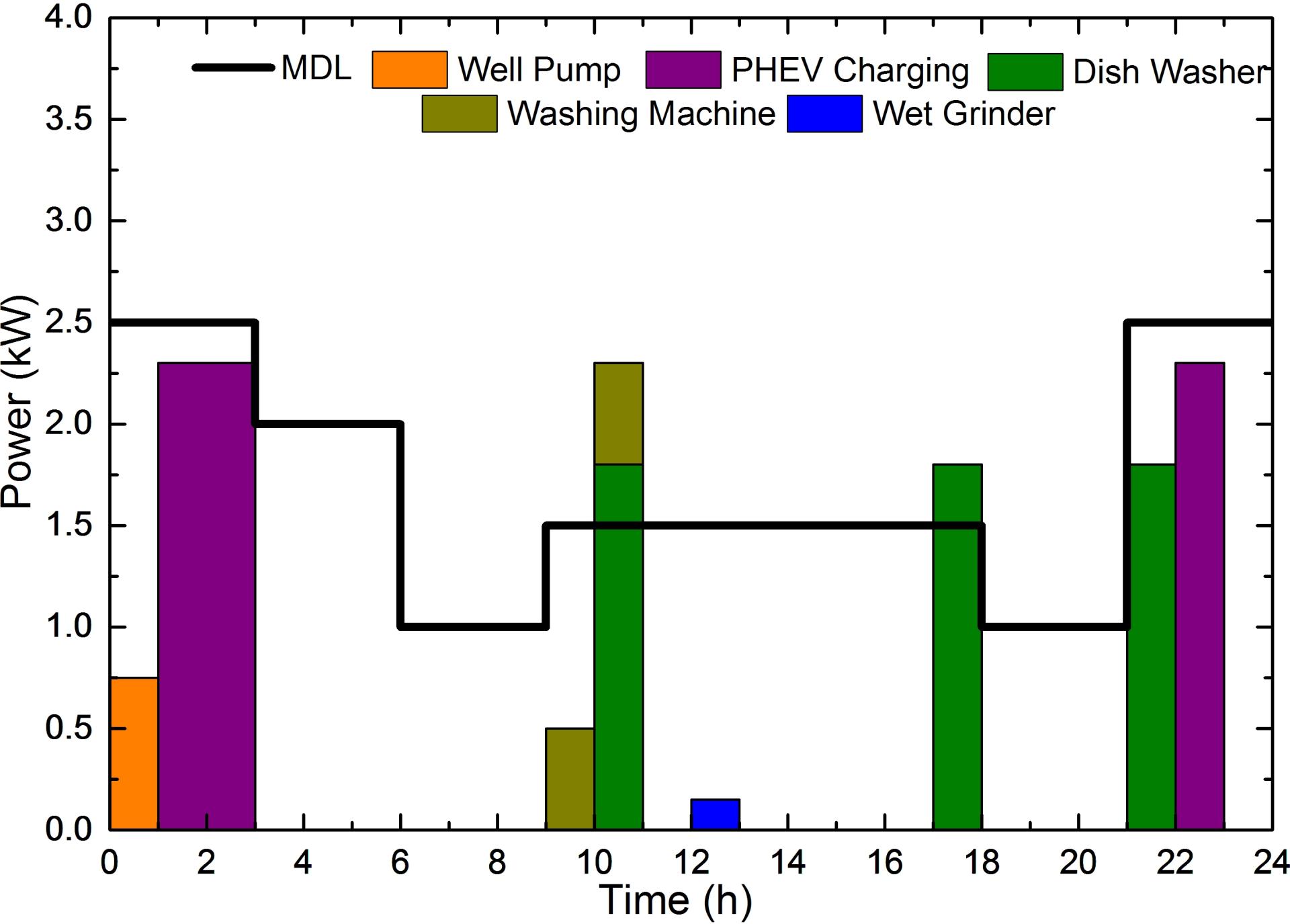}%
\label{6b}}

\caption{Consumer's load profile of a day}
\end{figure}

\textbf{Case I:} Consumer does not use any scheduling algorithm. All loads are scheduled by human intervention. Consumer's load profile for a day along with MDL provided by the utility is depicted in Figure 6a. 

\textbf{Case II:} In this case, the consumer uses SLNs to log the $\alpha$, $\beta$, $\gamma$ values for individual loads. Load scheduling does not involve any human intervention but is the result of the coordinated operation of SLNs and LMU in the developed HAN. Figure 6b represents consumer's load profile for a day using same MDL variation as in the case I. 

It can be observed from the results that the consumer's consumption in case II has crossed the MDL for less duration than in case I. Cost analysis of both the cases is shown in Table 1. 

\begin{table}[htbp]
\renewcommand{\arraystretch}{1.3}
\caption{Cost Analysis}
\label{loads}
\centering
\begin{tabular}{c c | c c}
\hline
\hline
\multicolumn{4}{l}{Penalty = $\$~x$ per kW--hr}\\
\hline
\multicolumn{2}{l|}{Without Scheduling Algorithm} & \multicolumn{2}{l}{With Scheduling Algorithm} \\
\hline
$E_1$ (kW--hr) & Penalty (\$) & $E_2$ (kW--hr) & Penalty (\$) \\
\hline
5.5 & $5.5~\times~x$ & 1 & $1~\times~x$ \\
\hline
\multicolumn{4}{l}{Savings $(E_1 - E_2)$ $=$ $4.5 \times x$ \$ }\\
\hline
\hline
\multicolumn{4}{l}{$(E_1, E_2)$: Total energy consumption beyond MDL}\\
\end{tabular}
\end{table}

A considerable reduction in penalty cost is obtained in case II. However, the amount of reduction in cost completely depends upon the efficacy of the scheduling algorithm employed. The proposed SLN can be used along with any kind of scheduling algorithm running in LMU. A limitation observed during the case study is a 7--9 second delay in communication between LMU and SLN. This delay is practically insignificant in the operation of schedulable loads like PHEV charging, well pump, dishwasher, etc.

\section{Conclusion}
A low--cost smart load node proposed in this article is a feasible solution for intelligent operation of non-smart residential loads in smart grid environment. Appropriately orchestrated operation of smart load nodes along with load management unit has demonstrated the smart operation of non-smart home appliances. The implemented smart load node can act as wireless power and power factor meter for any kind of load. Incorporation of Wi--Fi transceiver has facilitated wireless transfer of the electrical parameters of the load as well as user inputs from distributed multiple smart load nodes to the load management unit. The preceded case study has demonstrated the implementation of load management algorithm for non--smart loads using smart load nodes under smart grid paradigm.

\section*{Acknowledgment}
This research work was supported by Ministry of Electronics and Information Technology (MeitY), Government of India under Visvesvaraya Young Faculty Research Fellowship (Grant No. PhD--MLA--4(16)/2014 dated 7th April 2016).

\section*{About the Authors}
\textbf{Shashank~Singh} (shashanksingh0110@gmail.com) is currently a Master's student at the Department of Electrical and Electronics Engineering, National Institute of Technology (NIT) Tiruchirappalli, Tamil Nadu, India. He earned his Bachelor of Technology degree with Honours in Electrical and Electronics Engineering.

\textbf{Amit Roy} (amitroy.790@gmail.com) earned his Master's degree in Power Systems from National Institute of Technology (NIT) Tiruchirappalli, Tamil Nadu, India.

\textbf{Selvan M.P.} (selvanmp@nitt.edu) is currently an Associate Professor with the Electrical and Electronics Engineering Department, National Institute of Technology Tiruchirappalli, India, where he is associated with the Hybrid Electrical Systems Laboratory. He is a member of IEEE. He holds a Ph.D. degree from Indian Institute of Technology (IIT) Madras. He has published more than 100 technical research papers in various national, international conferences and journals.

\bibliographystyle{IEEEtran}
\bibliography{Ref}

\end{document}